\documentclass{aa}  
\usepackage{multirow} 
\usepackage{graphicx}
\usepackage{txfonts}
\usepackage{lipsum}
\usepackage{subcaption}         
\usepackage{lscape}       
\usepackage{hyperref}
\usepackage{placeins}           
\newcommand{\micron}{\ensuremath{\mu\mathrm{m}}}       

\begin{document}

\title{Effect of surface magnetic fields on limb darkening in main-sequence stars}

   \author{N. Kostogryz\inst{1}\fnmsep\thanks{Corresponding author: kostogryz@mps.mpg.de}
        \and A. I. Shapiro\inst{2, 1}
        \and V. Witzke\inst{2}
        \and T. Bhatia\inst{1}
        \and S. K. Solanki\inst{1, 3}
        \and I. Kuhlemann\inst{1}
        \and V. Vasilyev\inst{1}
        \and Y. C. Unruh\inst{4}
        }

   \institute{Max Planck Institute for Solar System Research,
Justus-von-Liebig-Weg 3, 37077 G\"ottingen, Germany
   \and University of Graz, Institute of Physics, Universit\"atsplatz 5, 8010 Graz, Austria
   \and School of Space Research, Kyung Hee University, Yongin, Gyeonggi 17104, Republic of Korea
   \and Department of Physics, Imperial College London, London SW7 2AZ, UK}
   \date{}

  \abstract
   {Stellar limb darkening encodes the thermal and radiative structure of stellar photospheres and is a key ingredient in modeling transit light curves and transmission spectra. It was recently shown that stellar surface magnetic fields modify limb darkening in stars with near-solar fundamental parameters, and that only magnetic models can reproduce high-precision transit observations for such stars. However, for stars with non-solar fundamental parameters, the magnitude of the magnetic effect on limb darkening remain unconstrained.}
   {We aim to investigate how surface magnetism affects stellar limb darkening across a range of fundamental parameters and to provide the community with center-to-limb spectra of stars at different magnetization levels.} 
   {We use the MPS-ATLAS code to compute synthetic spectra from 3D radiative magnetohydrodynamic box-in-a-star simulations performed with the MURaM code. These simulations self-consistently capture photospheric magneto-convection without relying on ad hoc parameterizations. We perform calculations for main-sequence stars at solar metallicity with effective temperatures in the range $T_{\mathrm{eff}} = 3200$–$6800$ K. For stars with solar effective temperature we also consider metal-poor, $\mathrm{[M/H]} = -1.0$, and metal-rich, $\mathrm{[M/H]} = 0.5$, cases.}
   {We show that the magnitude of the magnetic effect depends strongly on stellar fundamental parameters, increasing toward hotter and more metal-rich stars. Overall, limb darkening is significantly affected by magnetic fields in K-, G-, and F-dwarfs, while the effect becomes negligible in M-dwarfs. We release a public database of synthetic spectra at 10 disk positions \href{https://doi.org/10.17617/3.FBTIYY}{https://doi.org/10.17617/3.FBTIYY}.
   }
   {}

   \maketitle
    \nolinenumbers

\section{Introduction} \label{sec:intro}
In visible light, stars appear darker towards their limb than at the center of the disk. This is because, near the limb, the line of sight samples higher and cooler layers of the stellar atmosphere. This phenomenon, known as stellar limb darkening, has been recognized for over a century, with foundational work by \cite{Schwarzschild_1906} and \cite{Milne_1921}. Remarkably, this classical problem was addressed even before the mechanisms behind stellar energy generation were understood by \cite{bethe1939}. For several decades thereafter, 
research on limb darkening was on a low gear with interest driven mainly by interpreting light curves of eclipsing binary stars \citep{Kopal_1950}, interferometry \citep{Hestroffer_1997}, and studies of stellar variability caused by their rotation \citep{Lanza2003}. 

With the advent of high-precision transit photometry and transmission spectroscopy, studies of stellar limb darkening has regained significant attention. Stellar limb darkening plays a crucial role in interpreting transit light curves, as it influences both the transit profile and depth. Consequently, it directly affects the determination of planetary radius \citep{EspinozaJordan2015, Csizmadia2018LimbDarkening, Keers2024} and atmospheric composition through transmission spectroscopy \citep{seager_sasselov_2000, brown_2001, charbonneau_2002}. As a result, numerous grids and models have since been developed to characterize limb darkening across a broad range of stellar parameters, including effective temperature ($T_{\rm{eff}}$), metallicity ($\rm{[M/H]}$), and surface gravity ($\log g$),  such as \texttt{ATLAS9} \citep{Kurucz2005}, \texttt{MARCS} \citep{Gustafsson2008}, \texttt{PHOENIX} \citep{Husser2013}, \texttt{STAGGER} \citep{Magic_2015A&A} and \texttt{MPS-ATLAS} \citep{kostogryz2022}. 

Despite such a substantial effort models have fallen short when trying to explain observations. Recently, \citet{PatelEspinoza2022} empirically inferred limb-darkening coefficients from TESS transit light curves and reported significant offsets with respect to theoretical predictions, especially for cooler stars. A similar trend was pointed out by \citet{maxted_2023}, who compared observed and modeled limb darkening for a selected sample of Kepler and TESS stars with well-determined stellar parameters hosting transiting exoplanets. They found that, although available models such as \texttt{MPS-ATLAS}, \texttt{ATLAS9}, \texttt{PHOENIX}, \texttt{STAGGER}, and \texttt{MARCS} are broadly consistent with each other, they all predict a steeper intensity drop toward the stellar limb than is observed. This inconsistency, often referred to as the {\it limb-darkening conundrum}, is also evident in other observations, including analysis of James Webb Space Telescope (JWST) transit light curves of WASP-39 \citep{rustamkulov_2023_JWST}, interferometric observations of Alpha Centauri obtained with the \texttt{VLTI/PIONIER} optical interferometer \citep{kervella_2017, vlti, pionier_2010}.

One line of work has approached the offset as a methodological problem, refining how theoretical limb-darkening coefficients are derived and compared to transit-derived values \citep{howarth_2011, morello_2020, PatelEspinoza2022}, or examining possible biases in the empirically inferred coefficients \citep{coulombe_2024}. These studies focused on the procedures used to compare observed and modeled limb darkening profiles.

A separate line of work instead focused on the physics of the limb darkening itself. \citet{ludwig23} and \citet{kostogryz_2024} identified a promising physical mechanism capable of explaining the disagreement between measurements and models.
They identified an important limitation of earlier limb-darkening models: they all assumed non-magnetic stellar atmospheres. At the same time, the atmospheres of all lower main-sequence stars are believed to be magnetized. First, a global dynamo that operates deep in the convective-zone  builds large-scale fields. Second, a near-surface small-scale turbulent dynamo (SSD) fills the photosphere of even the most inactive stars with mixed-polarity magnetic fields \citep{voegler_schluessler_2007, danilovic_2010, bhatia_2022, witzke_2024, bhatia_2024}. The SSD sets a baseline level of activity and its effect  on stellar atmospheric structure and emergent spectra. The global dynamo leads to the formation of magnetic features, such as spots and faculae (sometimes refereed to as plage or bright spots in the  exoplanetary community). Capturing the effects of small- and large-scale magnetic fields on limb darkening requires radiative magnetohydrodynamic (MHD) simulations that self-consistently model photospheric magneto-convection and its coupling to radiation. Such simulations, performed by \citet{ludwig23} using the \texttt{CO5BOLD} code \citep{freytag_2012} and by \citet{kostogryz_2024} using the \texttt{MURaM} code \citep{Voegler_2005, Rempel2014}, were carried out for a solar twin (i.e., a star with solar effective temperature, surface gravity, and elemental composition). They showed that including surface magnetic fields is essential for reconciling limb-darkening calculations with observations of stars having near-solar parameters \citep[see the Methods section of][for a discussion of the domain of applicability]{kostogryz_2024}. The main effect was produced by faculae and magnetic network that become increasingly brighter (relative to non-magnetic stellar surface) towards the limb and flatten the center-to-limb intensity variation.

The calculations of \citet{ludwig23} and \citet{kostogryz_2024} were performed for a solar twin and are not directly applicable to stars with substantially different fundamental parameters. Indeed, 3D radiative MHD simulations with the \texttt{MURaM} code predict that facular and network contrasts depend on stellar parameters, generally decreasing toward cooler effective temperatures \citep{Beeck2015A&Athird, Salhab2018, Norris_2023, Shapiro2026, bhatia_2026} and toward lower metallicities \citep{Shapiro2026}.
This dependence reflects a competition between geometric/radiative brightening due to the hot-wall effect in Wilson-depressed magnetic flux concentrations and magnetic suppression of the convection energy flux \citep[see, e.g. Figure 5 and its discussion in][]{Solanki_et_al_2013}. As near-surface temperature gradients and radiative heating efficiency change across the HR diagram, the facular contrast even flips its sign, so for sufficiently cool stars faculae may appear darker than the non-magnetic surface rather than brighter \citep{Shapiro2026}. 

In this paper, we extend the calculations of \citet{kostogryz_2024} to stars with non-solar fundamental parameters and investigate how magnetic fields affect stellar limb darkening across a range of effective temperatures ($T_{\rm eff}$) and metallicities ($\rm M/H$). We publish stellar synthetic spectra as functions of limb position and surface magnetic fields for solar metallicity $\rm{M/H} = 0.0$ over $T_{\rm{eff}}$ corresponding to spectral types F3-M4, and for solar $T_{\rm{eff}}$ at three metallicities spanning -1.0 to +0.5. The full dataset is publicly available for the community use \href{https://doi.org/10.17617/3.FBTIYY}{https://doi.org/10.17617/3.FBTIYY}. 

\section{Grid of stellar spectra}\label{sec:grid}

The workflow of spectra computations proceeds in two steps: to compute quiet and magnetized stellar atmospheres of stars with different fundamental parameters, we first employ the radiative 3D MHD code \texttt{MURaM} \citep[which stands for {\bf M}PS/{\bf U}niversity of Chicago {\bf Ra}diative {\bf M}HD, see][]{Voegler_2005, Rempel2014}, and then synthesize the emergent spectra from these atmospheres using \texttt{MPS-ATLAS} \citep[{\bf M}erged {\bf P}arallelized {\bf S}implified ATLAS, see][]{mps-atlas_2021}.

\subsection{Model simulations with \texttt{MURaM}}\label{sec:muram_simulations}
We simulate stellar atmospheres with the 3D MHD \texttt{MURaM} code \citep{Voegler_2005,Rempel2014, witzke_2024}, which solves the conservative form of the magnetohydrodynamic (MHD) equations for a compressible, partially ionized plasma. The simulations are performed in the box-in-a-star regime, i.e. in a Cartesian box representing a small patch around the optical surface, spanning from the near‑surface convection zone to the top of the photosphere, including the temperature minimum. We adopt the F3, K4, and M0 simulations from \citet{bhatia_2026}, the solar-metallicity G2 simulation ($\mathrm{[M/H]} = 0.0$) from \citet{kostogryz_2024}, and the M4 simulation from \citet{Shapiro2026}. We further continued the G2 simulations at $\mathrm{[M/H]} = -1.0$ and $+0.5$ from \citet{witzke2022_SSD}, and complemented the existing set with new K0 and M2 simulations with solar metallicity. The provenance of the simulations is summarized in Table~\ref{tab:sim_origin}. Domain sizes, grid resolutions, and stellar parameters of all simulations are summarized in Table~\ref{tab:setup}.

To enable a direct comparison all these simulations followed a homogeneous setup. Namely, periodic boundary conditions were applied horizontally ($x$–$y$), the top boundary was open to outflows, but closed to inflows; the bottom boundary was open to both mass and magnetic flux. The horizontal extent of each computational domain was scaled with the local vertical pressure scale height — which sets the characteristic granulation size \citep{Magic_2013A&A} — and chosen to include roughly a dozen granules. The equation of state look-up table is generated using the \texttt{FreeEOS} code \citep{Irwin2012}. For the opacity tables we used the elemental composition from \cite{Asplund2009}. The radiative transfer in our simulations was computed using a multi‑group (opacity‑bin) method \citep{nordlund_1982}.
We adjusted the binning according to effective temperature and metallicity (see Figure~\ref{fig:opacity_bins} and the brief description in Appendix A). Further details of the simulation setup are given by \citet{Rempel2014,bhatia_2022,witzke_2024, bhatia_2026}. 

After the initial setup the hydrodynamic (HD) simulations were evolved until they reached a statistically steady (saturated) state. For the small-scale dynamo (SSD) cases, intended to represent quiet-star conditions, a random seed field with zero net flux was added to saturated HD simulations. For simulations representing active stars a vertical, unipolar magnetic field ($B_z$) was  added to the SSD setup. The magnetic simulations were run until the domain-averaged root-mean-square magnetic field saturated (which happens within several simulation box turnover times). 

All in  all, the suite comprises nine spectral types from F3 to M4. For each, we produced an HD run (no magnetic field), an SSD run (zero-net-flux dynamo, reproducing a quiet star), and three runs with an imposed vertical magnetic field (100, 200, and 300~G), representing faculae; see Table~\ref{tab:teff_simulations}.

All simulations used in this study have been run for a couple of stellar hours after reaching steady state to enable averaging over oscillatory and granulation effects in the resulting spectra (see Sect.~\ref{sec:mpsa_simulations})

\begin{table*}[]
    \centering
    \caption{ Setups of the \texttt{MURaM} simulation for stars with their fundamental parameters.} 
    \begin{tabular}{c|c|c|c|c|c|c|c}
        Star & $\rm{M/H}$ & T$_{\rm{eff}}$, K & $\log g$, cm s$^{-2}$ & $h$, pixels & $x$, pixels & $\Delta h$, km & $\Delta x$, km \\
        \hline
        F3 & 0.0 & 6724.8 & 4.0 & 1000 & 512 & 13 & 45 \\
        G2 & 0.0 & 5783.8 & 4.438 & 500 & 512 & 10 & 17.5 \\
        G2-mp & -1.0 & 5753.1 & 4.438 & 500 & 512 & 10 & 17.5 \\
        G2-mr & 0.5 & 5776.9 & 4.438 & 500 & 512 & 10 & 17.5 \\
        K0 & 0.0 & 5257.2 & 4.4& 480 & 512 & 10 & 17.5 \\
        K4 & 0.0 & 4592.7 & 4.609 & 500 & 512 & 4.6 & 8.2\\
        M0 & 0.0 & 3752.2 & 4.826 & 500 & 512 & 2.3 & 4.0\\
        M2 & 0.0 & 3531.8 & 5.0 & 500 & 512 & 1.1 & 2.0 \\
        M4 & 0.0 & 3219.2 & 5.0 & 500 & 512 & 0.7 & 1.2 
    \end{tabular}
    \tablefoot{Effective temperatures in the table correspond to quiet stars (SSD), represented by \texttt{MURaM} small-scale dynamo simulations.
    The table also reports the domain sizes and grid resolutions: $h$ and $\Delta h$ denote the vertical extent of the box (pixels) and the vertical grid spacing (km), respectively; $x$ and $\Delta x$ denote the horizontal extent (both $x$ and $y$; pixels) and the horizontal grid spacing (km), respectively. The G2 model corresponds to our solar case; G2-mp denotes a metal-poor case with $[\mathrm{M}/\mathrm{H}]=-1.0$, and G2-mr denotes a metal-rich case with $[\mathrm{M}/\mathrm{H}]=+0.5$.}
    \label{tab:setup}
\end{table*}

\subsection{Spectral synthesis with \texttt{MPS-ATLAS}}
\label{sec:mpsa_simulations}

For each stellar model and magnetization level we extract a time series of snapshots from the thermally relaxed phase at a cadence of 90\,s; the total number of snapshots per model is listed in
Table~\ref{tab:teff_simulations}. For each snapshot we compute emergent intensities over wavelengths from 200 to 10\,000\,nm and for ten viewing angles $\mu = \cos\theta$, from disk center ($\mu = 1$) to near the limb in steps of $\Delta\mu = 0.1$.

To compute emergent spectra from the 3D simulations we use the latest version of the \texttt{MPS-ATLAS} code. The radiative transfer is solved in LTE using a ray-by-ray scheme: for each simulation snapshot the code follows a large number of parallel 1D rays through the 3D cube. Spectra at the disk center of the stars correspond to vertical rays, and spectra at other viewing angles are calculated from inclined rays with temperature, pressure, and density being interpolated onto each of the rays. For a given $\mu$ and time the emergent specific intensity is obtained by averaging over all corresponding rays, and $I_\lambda(\mu)$ is finally computed as a temporal average. 

\texttt{MPS-ATLAS} relies on pre-tabulated opacities as a function of thermodynamic parameters and composition. In the present work we employ opacity-distribution-function (ODF) tables based on the classical ODF approach \citep{HubenMihalas} and its generalized implementation in \texttt{MPS-ATLAS} \citep{cernetic_odf_2019, Anusha_odf_2021}. These tables are constructed by rebinning high-resolution opacities computed at $R=500000$ onto a lower-resolution wavelength grid ($R \approx 400$ in the visible spectral domain), while accounting for more than 100 million atomic and molecular transitions \citep{Kurucz_2005_Atlas12_9}. Such low-resolution ODF tables are well suited for the calculation of broadband spectra and center-to-limb variations. The opacity tables used within \texttt{MURaM} are generated with the same \texttt{MPS-ATLAS} setup, ensuring that the opacities in the 3D radiative--(magneto)hydrodynamic calculations and in the subsequent spectral synthesis are fully consistent.

For each simulation, we compute two values of the effective temperature. One value (third column in Table~\ref{tab:teff_simulations}) is derived from integrated synthetic spectra computed with the \texttt{MPS-ATLAS} code using ODFs. This value represents the effective temperature of a star corresponding to the spectra we provide.
The second value (fourth column in Table~\ref{tab:teff_simulations}) is derived from the total radiative flux computed with \texttt{MURaM}, which relies on a less detailed multi-group radiative transfer scheme than that used in \texttt{MPS-ATLAS}. In contrast to the value derived from spectra, it represents the internal MURaM value that defines the temperature structure of our simulations.
The two values are therefore not identical: more accurate radiative transfer in MURaM brings them closer together, at the cost of more time-consuming simulations. The agreement between the two temperature estimates in Table~\ref{tab:teff_simulations}, typically at the percent level, indicates that the adopted opacity-bin prescriptions provide a reasonable representation of the radiative energy balance for the present simulations. Nevertheless, the remaining differences reflect the approximate nature of the multi-group opacity treatment, and further optimization of the binning is left for future work.

Granulation and oscillations lead to cube-to-cube temperature fluctuations with a root-mean-square amplitude of up to about 10~K (fifth column in Table~\ref{tab:teff_simulations}). The large number of simulation cubes used to compute the mean spectra (last column in Table~\ref{tab:teff_simulations}) implies that the uncertainty of the mean effective temperature is of order 1--2~K. 

\begin{table*}
    \centering
    \begin{tabular}{c|c|c|c|c|c}
        Star & Magnetization & $T_{\rm eff}$ from spectra [K] & $T_{\rm eff}$ [K] & $\sigma$ [K] & Number of cubes \\
        \hline
        \multirow{5}{*}{F3} & HD & 6806.7 & 6736.4 & 6.4 & 44\\
                                 & SSD & 6794.4 & 6724.8 & 7.9 & 76\\
                                 & 100~G  & 6837.6 & 6741.2 & 7.1 & 35\\
                                 & 200~G  & 6878.8 & 6749.0 & 9.8 & 50 \\
                                 & 300~G  & 6914.5 & 6750.4 & 11.1 & 20\\\hline
        \multirow{5}{*}{G2} & HD &5777.9 & 5780.1 & 9.1 & 44\\
                                 & SSD & 5791.6&5783.8 & 11.4 & 44\\
                                 & 100~G  & 5806.3&5808.1 & 7.1 & 48 \\
                                 & 200~G  & 5818.6&5822.6 & 7.1 & 42\\
                                 & 300~G  & 5816.7&5830.2 & 10.7 & 48\\\hline
        \multirow{5}{*}{G2-mp} & HD & 5749.9 & 5749.0 & 4.6 & 114 \\
                                 & SSD & 5754.7 &5753.1 & 4.5 & 136\\
                                 & 100~G  & 5764.9 &5762.9 & 5.9 & 79\\
                                 & 200~G  & 5771.6 & 5768.0 & 4.3 & 77\\
                                 & 300~G  & 5773.6&5768.3 & 4.2 & 76 \\\hline
        \multirow{5}{*}{G2-mr} & HD & 5780.5 & 5772.6 & 10.9 & 120 \\
                                 & SSD & 5785.4 &5776.9 & 12.3 & 66\\
                                 & 100~G  & 5817.0 & 5810.6 & 8.0 & 33\\
                                 & 200~G  & 5824.4 &5817.4 & 10.3 & 26\\
                                 & 300~G  & 5848.5 &5844.2 & 11.5 & 40\\\hline 
        \multirow{5}{*}{K0} & HD & 5239.7 &5243.9 & 7.7 & 103 \\
                                 & SSD & 5252.3 &5257.2 & 9.2 & 99\\
                                 & 100~G  & 5266.4&5272.3 & 7.2 & 90 \\
                                 & 200~G  & 5276.0&5283.5 & 8.0 & 38\\
                                 & 300~G  & 5285.8&5294.8 & 6.3 & 38\\\hline
        \multirow{5}{*}{K4} & HD & 4543.0 &4586.3 & 5.4 & 40 \\
                                 & SSD & 4549.8 &4592.7 & 5.9 & 39\\
                                 & 100~G  & 4561.4 &4603.8 & 6.4 &33 \\
                                 & 200~G  & 4566.7 &4609.1 & 4.6 & 41\\
                                 & 300~G  & 4576.6 &4618.3 & 4.2&42\\\hline
        \multirow{5}{*}{M0} & HD & 3707.3 &3750.8 & 1.1 & 121 \\
                                 & SSD & 3708.9 &3752.2 & 0.9 & 57\\
                                 & 100~G  & 3709.1 &3752.9 & 0.8 &29 \\
                                 & 200~G  & 3709.7 &3752.1 & 1.6 &28 \\
                                 & 300~G  & 3707.3 &3749.0 & 1.1 &32 \\\hline
        \multirow{5}{*}{M2}  & HD & 3518.3 & 3531.4 & 0.5 & 84  \\
                                 & SSD & 3513.5 & 3531.8 & 0.5 & 77\\
                                 & 100~G& 3511.3 & 3530.5 & 0.5 & 58\\
                                 & 200~G& 3507.0 & 3526.3 & 0.7 & 41\\
                                 & 300~G& 3502.4 & 3521.1& 0.6 & 37 \\\hline
        \multirow{5}{*}{M4} & HD & 3196.3 & 3219.0 & 0.8 & 213 \\
                                 & SSD & 3196.4 &3219.2 & 0.5 & 204 \\
                                 & 100~G  & 3194.9 &3217.2 & 0.5 & 320\\
                                 & 200~G  & 3191.0 & 3213.5& 0.8 & 130\\
                                 & 300~G  & 3187.8 &3210.5& 0.6 & 109\\\hline
    \end{tabular}
    \caption{{\bf Stellar parameters of the simulated models.} For each spectral type and magnetization level we list the mean effective temperature derived from the synthetic spectra ($T_{\rm eff}$ from spectra), the mean effective temperature of the underlying \texttt{MURaM} simulations obtained from the emergent bolometric intensity via the Stefan–Boltzmann law ($T_{\rm eff}$), the temporal standard deviation $\sigma$ of the latter, and the number of independent 3D cubes used in the spectral synthesis.} 
    \label{tab:teff_simulations}
\end{table*}
 
\section{Spectra and limb darkening of quiet stars}
In this section, we focus on spectra and limb darkening of quiet stars represented by our SSD setup.

As a first step, we examine intensity maps at a single diagnostic wavelength, $\lambda = 1.6\,\micron$, which lies close to the opacity continuum minimum in the near infrared. At this wavelength the contribution from spectral lines is small so that the radiation forms at comparatively large geometrical depths corresponding to the deepest layers of stellar atmospheres with the most vigorous convection. 

\begin{figure*}
    \centering
    \includegraphics[width=0.95\linewidth]{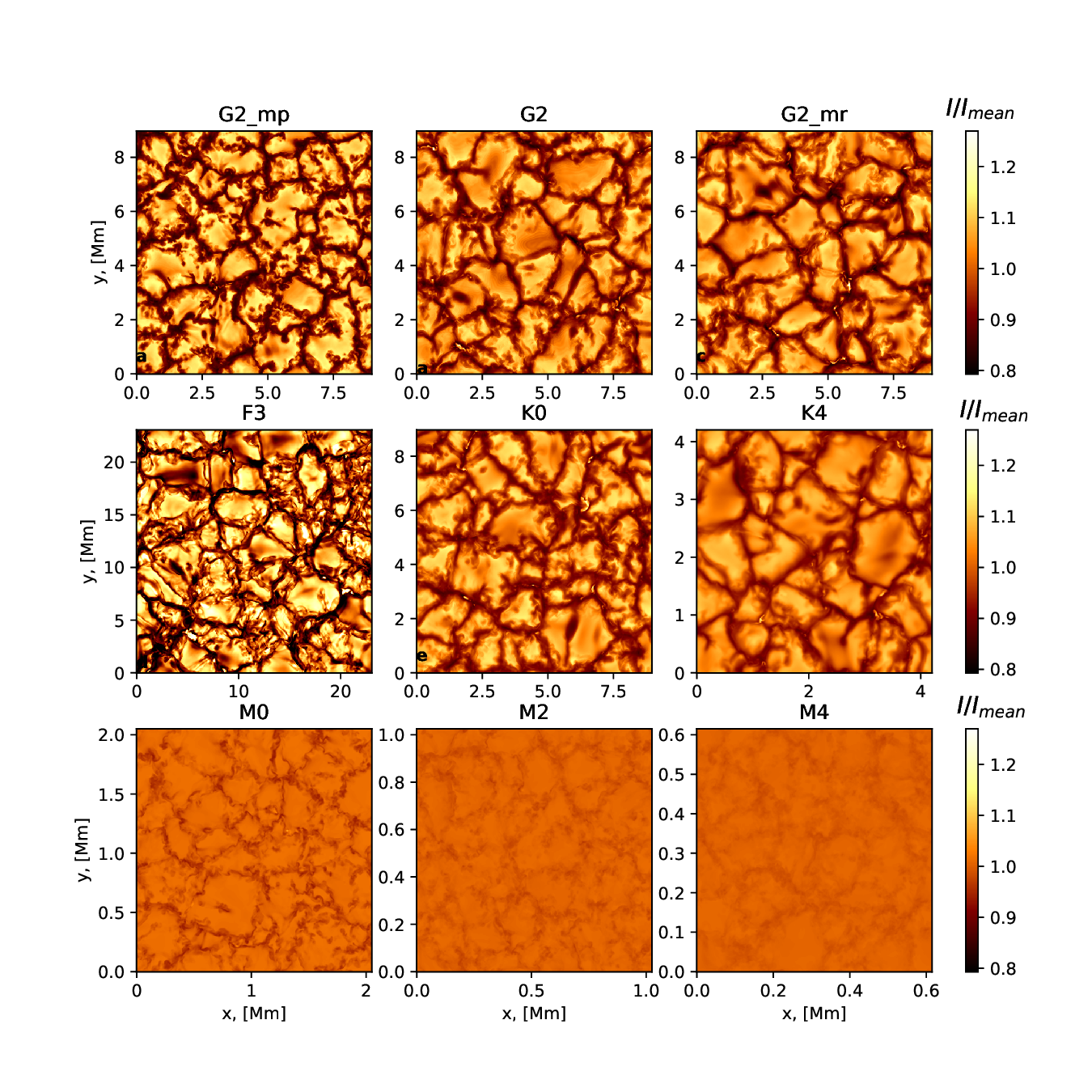}
    \caption{Emergent intensity images of quiet star models at $\lambda=1.6\,\micron$ (near the near-infrared opacity minimum) at disk center ($\mu=1$). Each panel shows a snapshot for one model, with the intensity $I$ normalized to the horizontal mean $\langle I\rangle$ (color bar on the right). From top left to bottom right the panels correspond to the following models: G2$_\mathrm{mp}$, G2, G2$_\mathrm{mr}$, F3, K0, K4, M0, M2, and M4 (see Table~\ref{tab:setup}). 
    }
    \label{fig:quiet_stars_image}
\end{figure*}

The corresponding emergent-intensity maps are shown in Figure~\ref{fig:quiet_stars_image}. The granulation pattern in the F--G models reveals pronounced structure and sharp intergranular lanes with high intensity contrast. Towards cooler K and especially M dwarfs, the pattern becomes progressively washed out. We highlight this trend in Figure~\ref{fig:quiet_stars_image} by using the same color scale. 

This trend is driven by two factors. First, temperature fluctuations decrease toward cooler stars as convection becomes less vigorous. Second, the sub-photospheric peak of the temperature fluctuations shifts to deeper layers, marking the transition from ``naked'' granulation in F- and G-dwarfs to ``hidden'' granulation in K- and M-dwarfs \citep[]{Nordlund_Dravins_1990, Beeck2013}.

\begin{figure*}
    \centering
    \includegraphics[width=0.99\linewidth]{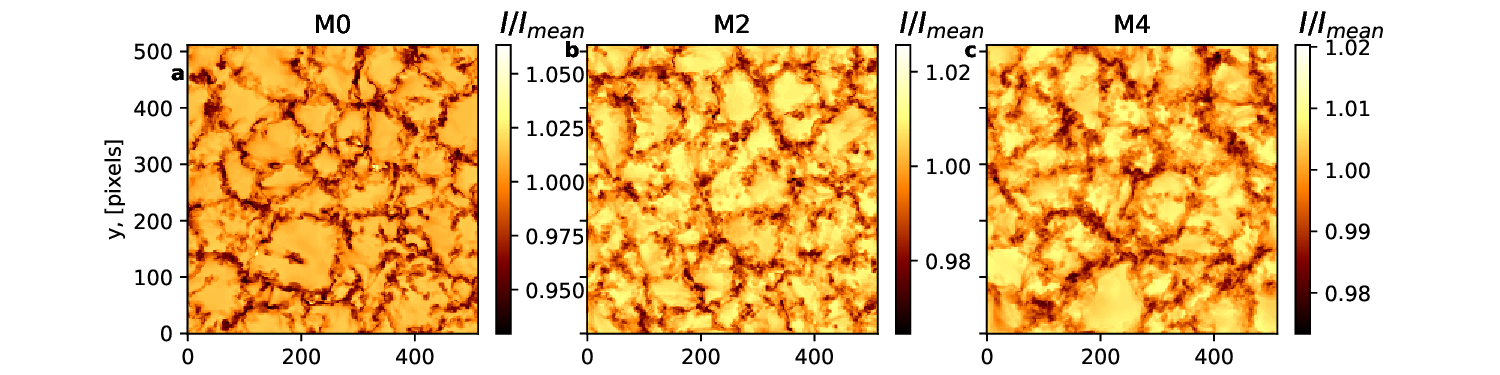}
    \caption{Emergent intensity at $\lambda = 1.6\,\micron$ at disk center for the quiet M-dwarf models M0, M2, and M4 (left to right). The intensity $I$ is normalized to the horizontal mean $\langle I\rangle$ in each snapshot, as indicated by the color bars. To emphasize the granulation pattern, the  dynamic range of the color scale is adjusted separately for each panel.}
    \label{fig:quiet_m_stars_image}
\end{figure*}

To better illustrate the weak granulation pattern in M-dwarfs, we show disk-center intensity maps in Figure~\ref{fig:quiet_m_stars_image}, with the color scale adjusted separately for each panel. Compared to hotter stars, the granulation pattern in M-dwarfs is less structured, consistent with the granulation being ``hidden'', where convection is most vigorous below the surface and only remnants are visible in the emergent intensity.

\begin{figure*}
    \centering
    \includegraphics[width=0.9\linewidth]{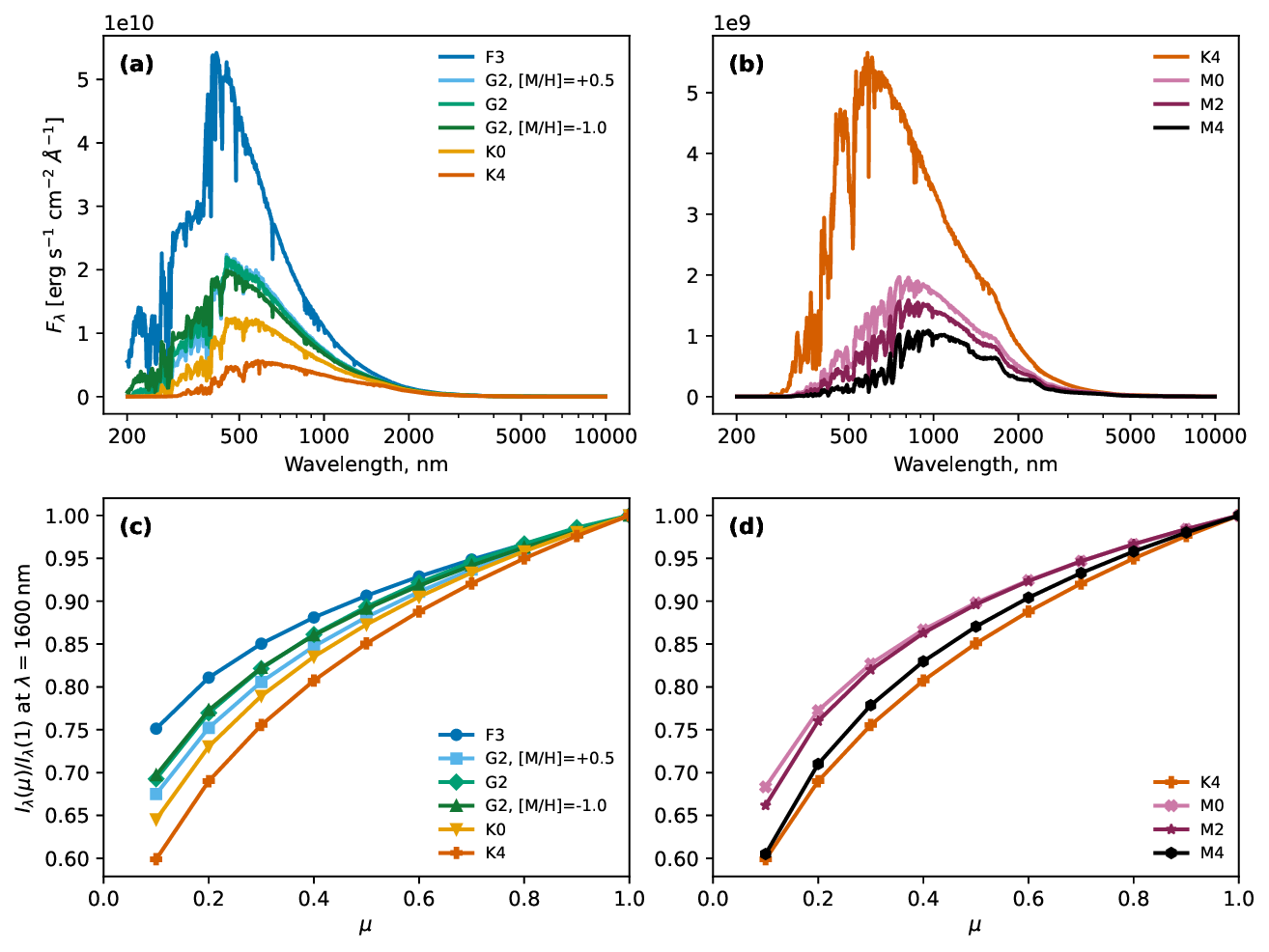}
    \caption{{\bf Spectra and limb darkening of quiet stars.}
\textit{Top panels:} Disk-integrated flux $F_\lambda$ as a function of wavelength for the small-scale dynamo reference simulations. The left panel shows the F3, G2-mr, G2, G2-mp, K0, and K4 models, while the right panel shows the M0, M2, and M4 models together with K4 for comparison. The two upper panels use independent y-axis scalings. \textit{Bottom panels:} Center-to-limb variation of the intensity at $\lambda = 1.6,\micron$ (opacity minimum), with $I_\lambda(\mu)$ normalized to the disk-center value $I_\lambda(\mu=1)$. The left panel displays the F--K models, while the right panel shows the M-dwarf models with K4 repeated to facilitate comparison.}
    \label{fig:spectra_ld_16}
\end{figure*}

We now quantify the radiative properties of our quiet-star models in terms of their spectral energy distributions (SEDs) and limb darkening. Our results for SEDs of quiet stars are consistent with previous models, including the 1D MPS-ATLAS calculations of \cite{kostogryz2022}. The spectral energy distributions exhibit the textbook behavior with effective temperature, namely a redward shift of the SED peak, a decreasing Balmer jump amplitude, and an overall reduction of the emergent flux as $T_{\rm eff}$ decreases (Fig.~\ref{fig:spectra_ld_16}). Differences between the metal-poor and metal-rich cases highlight the effect of line blanketing: with increasing metallicity, the opacity rises and suppresses the emergent flux at short wavelengths. The blocked radiation is redistributed toward longer wavelengths, leading to enhanced flux in the red and near-infrared. Each model produces a monotonic limb-darkening trend, but with distinct profile shapes across viewing angles. 

Limb darkening becomes progressively stronger from F to K stars (left bottom panel of Figure~\ref{fig:spectra_ld_16}. Interestingly, the M-dwarf models, however, do not form a simple monotonic continuation of this trend. The M0 and M2 models exhibit flatter center-to-limb variations than K4 between disk center and intermediate viewing angles, up to about $\mu \approx 0.4$ (see, bottom right panel of Figure~\ref{fig:spectra_ld_16}). This behavior is consistent with the hidden-granulation phenomenon: near the visible surface, the energy is mainly transported by radiation, and the granulation pattern becomes less distinct (see Fig.~\ref{fig:quiet_m_stars_image}), implying a smaller temperature gradient than in hotter stars. Closer to the limb, however, the M-dwarf profiles steepen, and the intensity drops more rapidly with decreasing $\mu$ than in the K4 model. This is likely related to the increasing molecular opacity in the atmospheres of cool M dwarfs, which affects the formation height of the emergent radiation and can lead to steeper temperature gradients in the upper photosphere. As a result, the M-dwarf limb darkening is relatively flat over most of the disk but becomes steeper close to the limb.

Overall, these intensity maps and angle-dependent spectra provide a coherent baseline across our quiet-star grid from F3 to M4. In the following sections, we use this baseline as a reference to quantify the impact of magnetic fields on stellar spectra and limb darkening.

\section{Magnetic effect on stellar limb darkening}

We now turn to simulations of faculae and magnetic network. These are obtained by imposing a vertical magnetic field on SSD simulations (after they have reached a statistically steady state) and allowing the field to evolve for several convective turnover times (typically a few hours). During this time the initially uniform magnetic field is advected by convection into patches with field strengths, corresponding to mechanical equilibrium where magnetic and convective kinetic energies are comparable (about 500–600 G at the surface in the solar case), which very quickly collapse into flux tubes with predominantly vertical fields of 1.5–2 kG at the surface \citep[see][for a detailed discussion of the convective collapse]{Parker1978, Spruit1979, Rajaguru2002}, largely independently of spectral type, metallicity, and the initially imposed field strength \citep{Shapiro2026}. The strength of the imposed vertical field primarily affects the fraction of the surface occupied by the flux tubes, thereby producing a progressively more faculae-dominated intensity pattern and a larger net facular contribution to the emergent intensity.

\begin{figure*}
    \centering
    \includegraphics[width=0.99\linewidth]{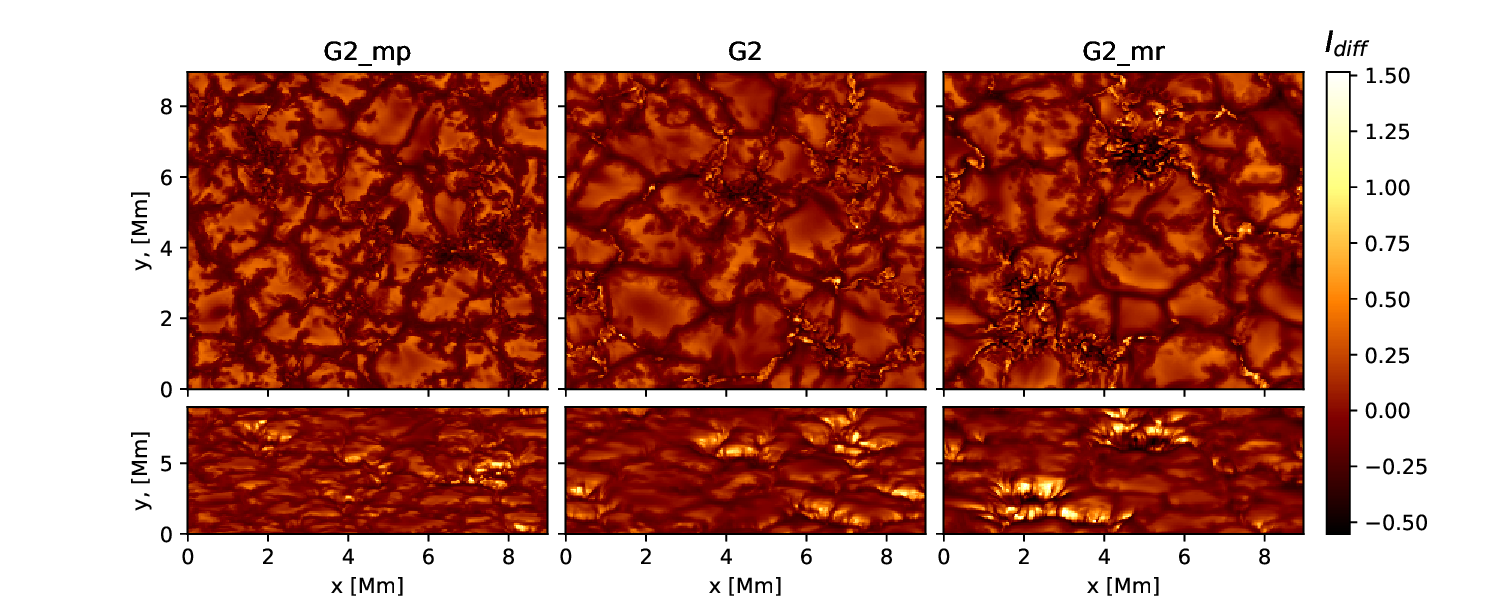}
    \includegraphics[width=0.99\linewidth]{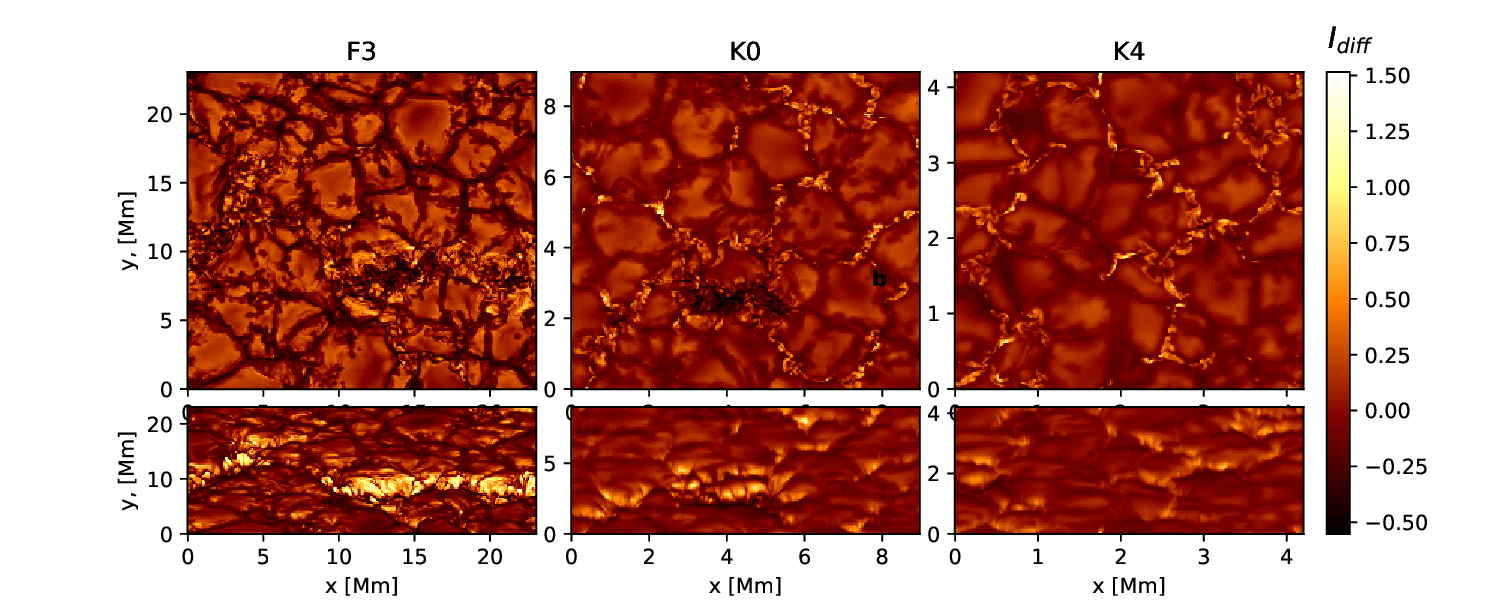}
    \includegraphics[width=0.99\linewidth]{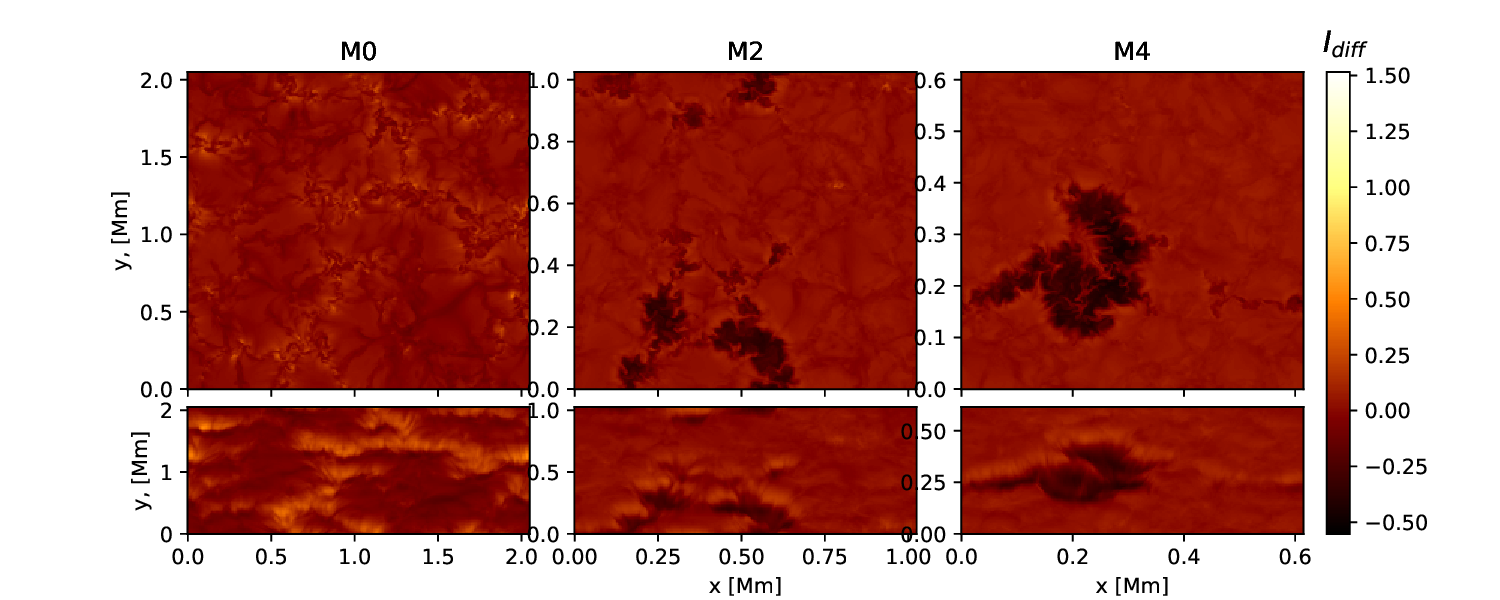}
    \caption{{\bf Facular intensity at $\lambda = 0.6\,\micron$ for the $B_z = 200$~G simulations.} Each pair of panels shows, for one spectral type, the relative difference, $I_{diff}$, between the spatially resolved intensity and its horizontal average. By construction, $I_{diff}=0$ marks the mean, while bright (dark) tones indicate regions brighter (darker) than average. From top to bottom:
    G2\_mp, G2, G2\_mr; F3, K0, K4; M0, M2, M4. For each star, the upper panel corresponds to disk center ($\mu = 1$)
    and the lower panel to an inclined view at $\mu = 0.4$.  In the
    hotter F--K models the limb views reveal extended bright facular patches associated with the hot walls of evacuated
    magnetic flux tubes, while the coolest M-dwarf models develop extended dark flux concentrations where strong fields
    inhibit convection. Among the G2 models, the facular brightening is strongest for the metal-rich case (G2\_mr) and
    weakest for the metal-poor case (G2\_mp), illustrating the role of metallicity.}
    \label{fig:facular_maps_06}
\end{figure*}

In what follows, we first focus on the $B_z = 200$~G runs and show how magnetic field  modifies the spatial intensity pattern, and then quantify the impact of different field strengths on the center-to-limb variation. Figure~\ref{fig:facular_maps_06} shows maps (at two viewing angles) of the intensity difference, $I_{diff}$ between the $B_z = 200$~G simulations and the horizontally averaged models ($I_{B=200G}/<I_{B=200G}> - 1$). Since magnetic effects are more pronounced in the visible wavelength range than at $1.6 \micron$ we show intensity maps calculated at $\lambda = 0.6\,\micron$.

The most striking feature of the disk-center maps for F--K dwarfs is that parts of the intergranular lanes exhibit pronounced brightening associated with the magnetic flux tubes formed in the integranular lanes by convective collapse.  Strong vertical fields partially evacuate the flux tubes, depressing the $\tau=1$ surface within them and exposing deeper, hotter layers of the surrounding plasma visible as the bright "hot walls" of the tube. This leads to the brightening (especially close to the limb when the hot walls are most visible). The brightening  occurs despite the inhibition of convection and the associated reduction of convective heating in magnetized regions \citep[][]{Rempel2011}. The brightening of the flux tubes weakens toward cooler and metal-poor dwarfs and is progressively replaced by darkening in the M2- and M4-dwarf simulations. In these stars, the near-surface atmospheric layers are  denser so that the evacuation of the flux tubes is weaker and, in contrast to hotter and metal-rich dwarfs, the deep layers remain hidden.

The limb view ($\mu = 0.4$) reveals bright facular patches that are even more pronounced than at disk center for the K-, G-, and F-type stars (Figure~\ref{fig:facular_maps_06}). This is not surprising, since hot walls are more clearly visible along inclined lines of sight \citep[][]{Spruit1976, Solanki1993, DePontieu2006, Solanki_et_al_2013}. 
In contrast, the dominant features in the M2 and M4 models are large, pore-like dark regions where strong magnetic flux concentrations suppress convection (Figure~\ref{fig:facular_maps_06}, bottom row). These dark structures more than compensate for the effect of the hot walls, which are significantly weakened (but still visible). 

\begin{figure*}
    \centering
    \includegraphics[width=0.9\linewidth]{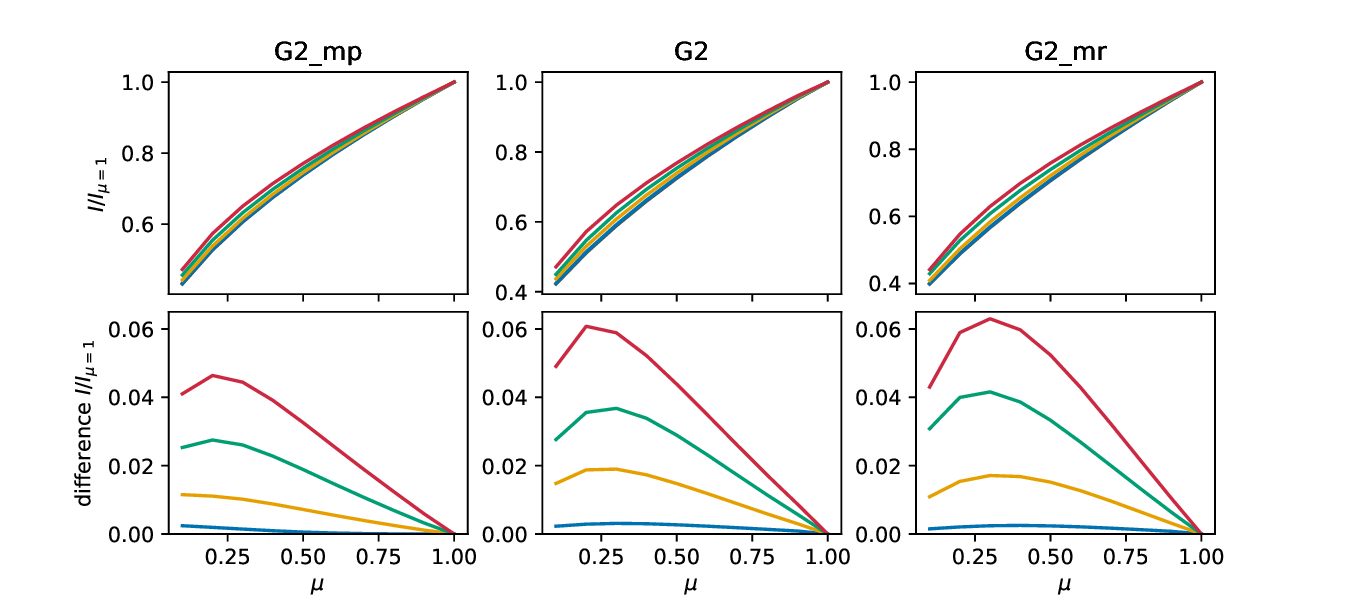}
    \includegraphics[width=0.9\linewidth]{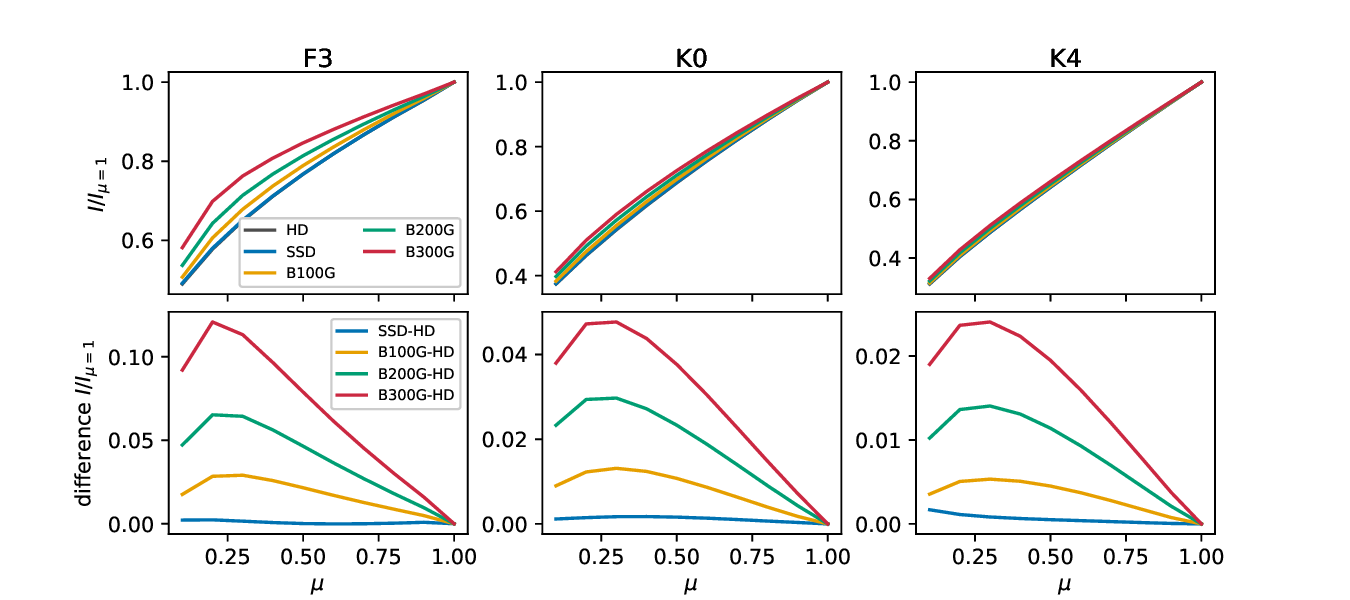}
    \includegraphics[width=0.9\linewidth]{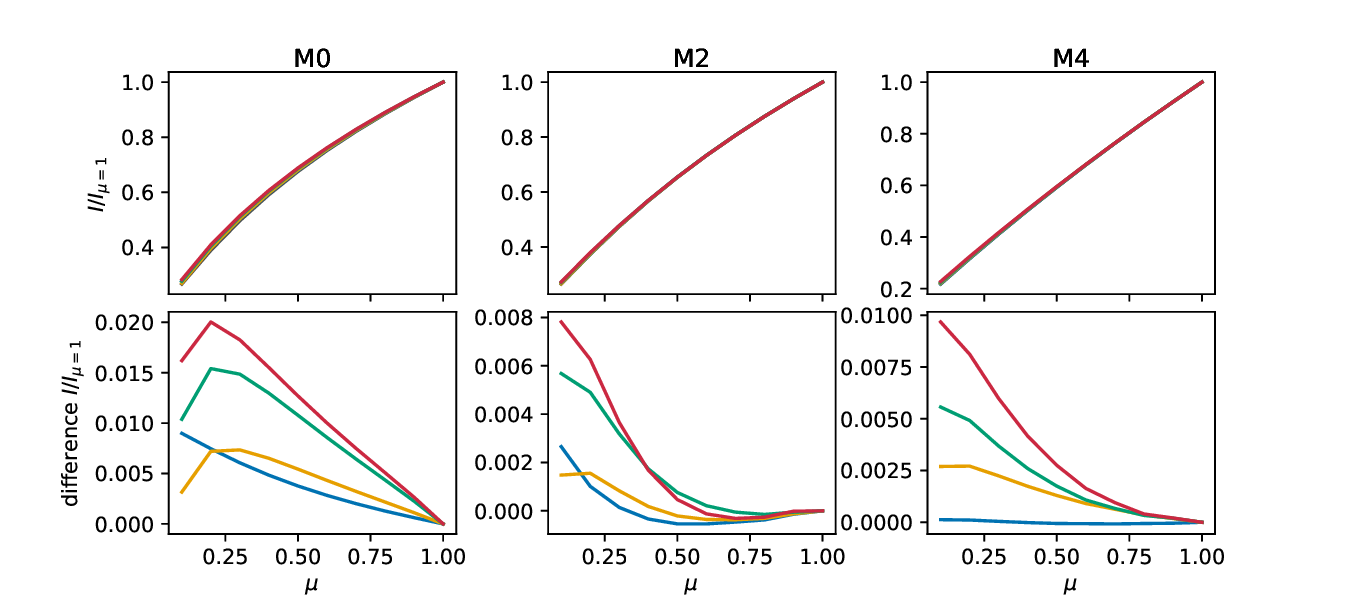}
    \caption{Effect of magnetic fields on limb darkening at
    $\lambda = 0.6\,\micron$. For each spectral type (columns), the upper panel shows the specific intensity as a function of viewing angle, normalized to the corresponding disk-center value, $I(\mu)/I(\mu_0)$  with $\mu_0 = 1$. The colors denote different magnetization levels: non-magnetic HD (black), SSD reference (blue), and imposed vertical fields of 100~G (orange), 200~G (green), and 300~G (red). The lower panels show the difference with respect to the SSD run, $\Delta I(\mu)$, normalized by the SSD disk-center intensity. Positive values indicate an intensity increase caused by the imposed field.}
    \label{fig:ld_faculae_06}
\end{figure*}

Introducing small-scale vertical fields systematically alters the center-to-limb variation by modifying the photospheric vertical temperature gradient and by introducing hot walls, whose brightness increases toward the limb  (Fig.~\ref{fig:ld_faculae_06}). 
This magnetic effect is especially strong in F–K dwarfs, where limb darkening becomes progressively weaker with increasing magnetization. Bright facular patches clearly visible in Figure~\ref{fig:facular_maps_06} produce a positive brightness excess near the limb (peaking around $\mu \simeq 0.2$--0.4). The amplitude of the effect decreases toward cooler stars, and M dwarfs show only a weak response of limb darkening to the magnetic field. Interestingly, the magnitude of the effect is metallicity dependent, as illustrated here for the G2 models: the reduction in limb darkening is strongest in the metal-rich case and weakest in the metal-poor case. We highlight the magnetic contribution to the limb darkening by plotting differences  between magnetic and non-magnetic limb darkening curves in lower panels of 
Fig.~\ref{fig:ld_faculae_06}.

Taken together, these results show that magnetization of order a few hundred gauss
can reshape the center-to-limb variations in F--K
dwarfs, while having only a modest effect on
cool M dwarfs, where magnetic suppression of convection dominates over the effect of hot walls and faculae appear locally dark. The strength of the magnetic effect on the limb darkening increases both with effective temperature and with metallicity. Our calculations thus provide a link between stellar parameters, photospheric magnetism, and the limb-darkening behavior.

\section{Stellar limb-darkening coefficients for the Kepler, TESS, and PLATO passbands}
\label{sec:ld_passbands}

Following \cite{maxted_2023}, we describe limb darkening in passbands of different instruments using the
transformed coefficients $(h'_1=I_{2/3}$, and $h'_2=I_{2/3}-I_{1/3}$, where $I_{2/3}$ and $I_{1/3}$ correspond to limb darkening at $\mu=2/3$  and $\mu=1/3$, respectively.  In this parameterization, $h'_1$
primarily traces the overall strength of limb darkening, while $h'_2$ captures
the curvature of the intensity profile. \citet{kostogryz_2024} computed the dependence of these coefficients on magnetic field for the Sun.
Here we extend their calculations to M4--F3 dwarfs with solar metallicity and G2-dwarfs with non-solar metallicity. We consider the passbands of Kepler \citep{borucki_2010}, TESS \citep{ricker_2015}, and PLATO \citep{rauer_2014}.

For each instrument, we compute band-integrated center-to-limb intensity profiles by convolving the monochromatic specific intensities with the instrumental response functions. Because CCD detectors count photons rather than measure radiant energy directly, we convert the intensity to photon counts,
\begin{equation}
I_{\rm pb}(\mu) = \int_{0}^{\infty} R_{\rm pb}(\lambda)\,
I(\lambda,\mu)\,\frac{\lambda}{hc}\,{\rm d}\lambda,
\end{equation}
where $R_{\rm pb}(\lambda)$ is the response function of passband ``pb'',
$h$ is Planck's constant, and $c$ is the
speed of light. 

\begin{figure*}
    \centering
    \includegraphics[width=0.9\linewidth]{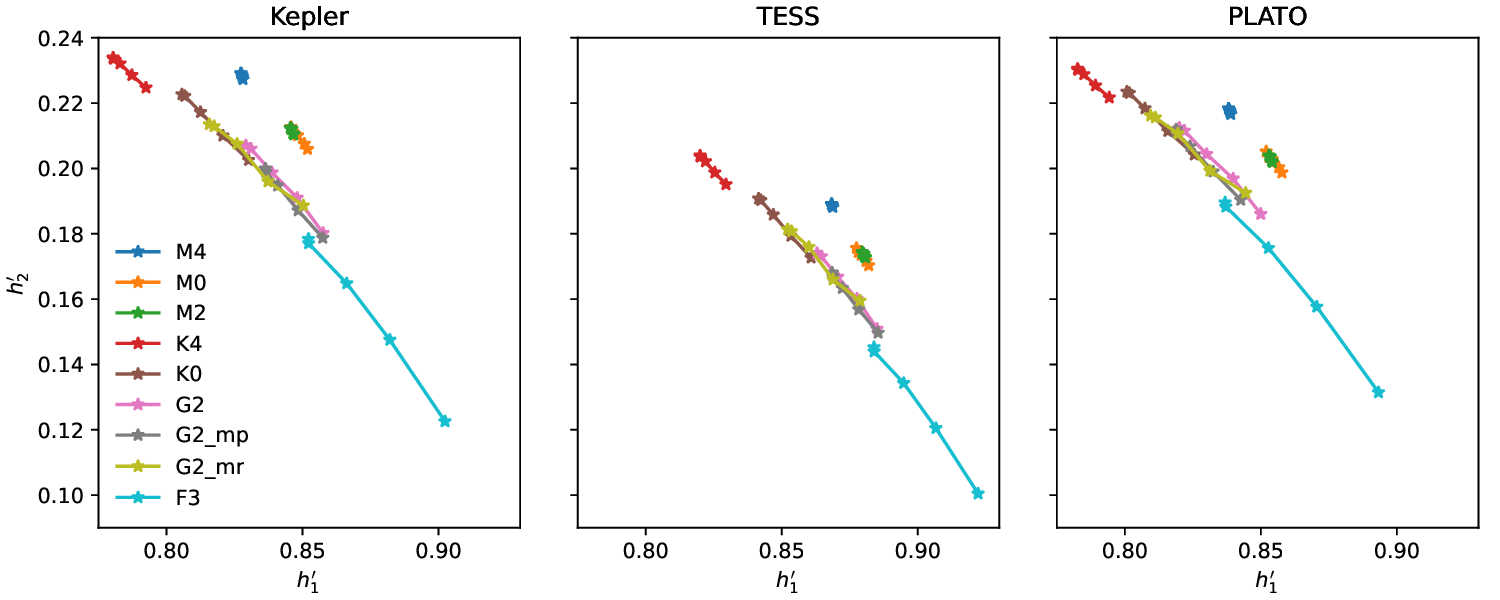}
    \caption{Stellar limb-darkening coefficients in the Kepler, TESS, and PLATO passbands, shown as transformed coefficients $h'_1=I_{2/3}$ and $h'_2=I_{2/3}-I_{1/3}$ \citep{maxted_2023, kostogryz_2024}. Colors indicate spectral type (F3–M4), and lines connect different magnetization levels (HD, SSD, 100–300~G) for each star. 
    }
    \label{fig:h1h2_missions}
\end{figure*}

The result is plotted in Figure~\ref{fig:h1h2_missions} for all the modeled stars considered here. Across all three passbands the models lie along a well-defined sequence in the
$(h'_1,h'_2)$ plane: hotter stars (F3) cluster at larger $h'_1$ and smaller
$h'_2$, whereas cooler stars (K and M dwarfs) are found toward lower $h'_1$ and
higher $h'_2$. This ordering reflects the systematic strengthening of limb
darkening toward cooler effective temperatures in these broad bands. Stars cluster along two distinct sequences: one formed by F--K dwarfs, and another by M-dwarfs. The separation between these sequences reflects differences in shapes between the limb darkening of M-dwarfs and hotter stars (see Figure~\ref{fig:spectra_ld_16}). Magnetization shifts points along each sequence. While HD and SSD are almost lying on top of each other, the 3 asterisks corresponding to 100, 200, 300 G simulations are well separated.  In line with the discussion above, the magnitude of the magnetic shift is strongly temperature dependent: it is largest for the hotter F- and G-type stars, smaller for K dwarfs, and almost negligible for M dwarfs. The shift is always in the same direction, toward larger $h_1'$ and smaller $h_2'$, indicating a systematic weakening of limb darkening. 

The limb darkening in Kepler and PLATO passbands are very similar, both for the field free and the magnetized atmospheres. By contrast, the limb darkening in TESS passband is systematically different, with larger $h_1'$ and smaller $h_2'$ than in Kepler and PLATO. This reflects the wavelength dependence of limb darkening, which weakens toward longer wavelengths. The magnetic effect is likewise wavelength dependent, decreasing in amplitude toward longer wavelengths.

\section{Conclusions}
\label{sec:conclusions}

Using 3D MHD models, we quantified how small-scale magnetic fields modify stellar surface brightness and limb darkening across solar-metallicity dwarfs from F to M (F3–M4) and G-dwarf models spanning multiple metallicities. We find that the magnetic impact on limb darkening increases with both effective temperature and metallicity: it is strongest in the hotter F-type models and in metal-rich G dwarfs, and it weakens toward lower [Fe/H] and cooler spectral types. In particular, the M-dwarf simulations develop predominantly dark magnetic structures and exhibit only weak changes in the global center-to-limb variation because these dark features have a similar center-to-limb variation as the quiet star.

A main deliverable of this work is a publicly available library of the low resolution stellar synthetic spectra as a function of limb position, magnetic field strength, and stellar parameters. In particular, the availability of spectra across different magnetization levels is essential for modeling transit contamination by stellar surface heterogeneity (e.g. faculae and magnetic network) and for quantifying variability signals driven by evolving surface magnetism. 

\section{Data availability}

Stellar spectra for all magnetization levels and viewing angles are available for community. We will issue new releases of the library as additional spectra become available. The underlying 3D simulation models for the stars considered in this work are available from the authors upon reasonable request.

\begin{acknowledgements}
NK, IK, and VV acknowledge support by German Aerospace Center (DLR) grants ``PLATO Data Center'' $50$OO$1501$ and $50$OP$1902$. AIS and VW was supported by the ERC Synergy Grant REVEAL under the European Union’s Horizon 2020 research and innovation program (grant no. 101118581). NK and AIS acknowledge support by Volkswagen Foundation (grant 9E126), TB and AIS acknowledge support from DFG grant SH1489/1. YCU was support by UKRI, grant ST/W000989/1. We acknowledge support by the Max Planck Computing and Data Facility.

\end{acknowledgements}

\bibliographystyle{aa} 
\bibliography{aa60518-26}

\clearpage
\begin{appendix}
\onecolumn
\section{Opacity binning in MURaM.}

\begin{figure*}[h!]
    \centering
    \includegraphics[width=0.9\linewidth]{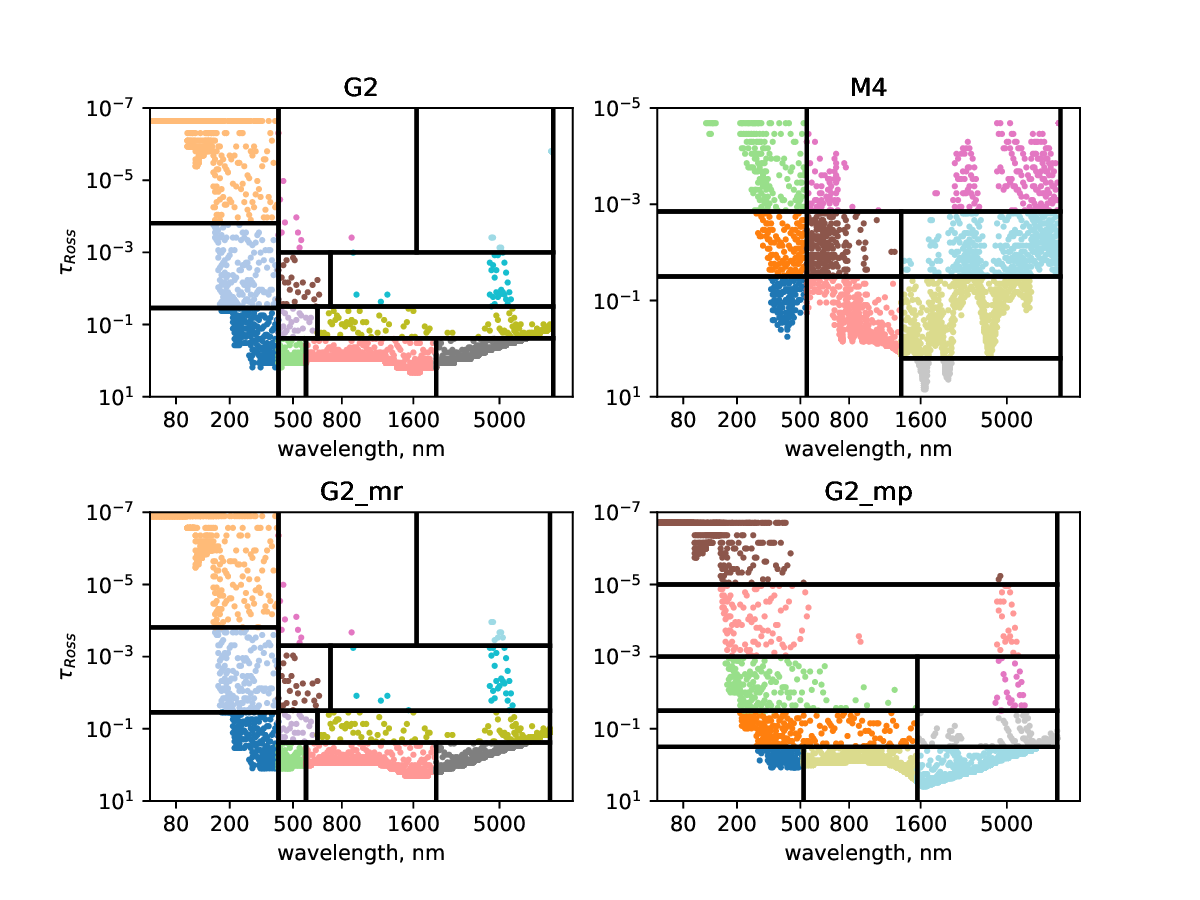}
    \caption{Opacity distributions and adopted multi-group thresholds for selected models. Colored points show individual opacity samples from the ODF tables mapped into the $(\lambda,\tau_{\mathrm{R}})$ plane (wavelength and Rosseland optical depth), with color indicating the assigned opacity group. Thick black lines mark the wavelength- and depth-dependent boundaries of the opacity bins used in the simulations. The three G2 panels (G2, G2\_mr, and G2\_mp) illustrate how the strong-line opacity pattern changes with metallicity and how the bin boundaries are adjusted accordingly. The M4 panel shows an adapted binning for the coolest models, reflecting the redistributed opacity in a cool, molecule-dominated atmosphere (representative of the M2/M4 binning).}
    \label{fig:opacity_bins}
\end{figure*}

We employ multi-group opacity binning, with bin boundaries defined in wavelength and Rosseland optical depth, to describe radiative opacities across the model grid. Different opacity-bin prescriptions are used depending on spectral type and composition. For the F3, K4, and M0 models, we adopt the opacity-bin thresholds used in the STAGGER simulations, as listed in Table~2 of \citet{beeck2012}. For the G2 models we use a modified binning scheme (illustrated in Fig.~\ref{fig:opacity_bins}, including the metallicity variants G2, G2\_mr, and G2\_mp), and for the coolest M dwarfs (M2 and M4) we adapt the binning to capture the redistributed opacity in molecule-dominated atmospheres; Fig.~\ref{fig:opacity_bins} shows M4 as a representative example.

To construct the modified/adapted schemes, we first compute a horizontally averaged mean atmospheric stratification from representative snapshots and establish the mapping to Rosseland optical depth, $\tau_{\rm R}$, using the corresponding Rosseland mean opacity. We then analyze the distribution of pre-tabulated opacity samples from the ODF tables in the $(\lambda,\tau_{\rm R})$ plane and place bin boundaries so that continuum windows and line-dominated regions are treated separately while the depth dependence of strong opacities is resolved by distinct bins in different atmospheric layers. The pre-tabulated opacity tables were updated several times to track the evolving time-averaged mean stratification. As shown in Fig.~\ref{fig:opacity_bins}, the three G2 panels highlight how the strong-line opacity pattern changes with metallicity and how the bin boundaries are adjusted accordingly, whereas the M4 panel exhibits a markedly different distribution reflecting the cooler, molecule-dominated atmosphere.

\section{Simulation runs with MURaM}
In this appendix, we summarize the \texttt{MURaM} simulation runs used in this work. Table~\ref{tab:sim_origin} lists the stellar models considered here and indicates, for each case, whether the corresponding simulation was taken from the literature, continued from a previously published calculation, or newly computed for this study.
\begin{table*}[]
    \centering
    \caption{Origin of the simulations used in this paper.} 
    \begin{tabular}{c|c|c|c}
        Star & New & Continued & Published\\
        \hline
        F3 & - & - & \cite{bhatia_2026} \\
        G2 & - & - & \cite{kostogryz_2024} \\
        G2-mr & - & \cite{witzke2022_SSD} & - \\
        G2-mp & - & \cite{witzke2022_SSD} & - \\
        K0 & this work & - & -\\
        K4 & - & - & \cite{bhatia_2026} \\
        M0 & - & - & \cite{bhatia_2026} \\
        M2 & this work & - & -\\
        M4 & - & - & \cite{Shapiro2026} 
    \end{tabular}
    \tablefoot{For each stellar model, we indicate whether the corresponding simulation is new in this work, continued from a previous calculation, or adopted from the published literature; the relevant references are given where applicable.}
    \label{tab:sim_origin}
\end{table*}

\end{appendix}
\twocolumn
\end{document}